\documentclass[sigconf]{acmart}

\usepackage{tabularx} % For tables that stretch to fit the width

\usepackage{enumitem}

\AtBeginDocument{%
  }

\setcopyright{acmlicensed}
\copyrightyear{2024}
\acmYear{2024}
\acmDOI{XXXXXXX.XXXXXXX}

\acmConference[ACM ICAIF '24]{ACM International Conference on AI in Finance}{Nov. 14--15, 2025}{Brooklyn, NY}

\begin{document}

\title{Revisiting Ensemble Methods for Stock Trading and Crypto Trading Tasks at ACM ICAIF FinRL Contests 2023/2024}

% \author{Anonymous}
 \author{Nikolaus Holzer, Keyi Wang}
 \authornote{Both authors contributed equally to this research.}
 \email{{nh2677,kw2914}@columbia.edu}
 \affiliation{%
   \institution{Columbia University}
   \city{New York}
   \state{New York}
   \country{USA}
 }

 \author{Kairong Xiao}
 \email{kx2139@columbia.edu }
 \affiliation{%
   \institution{Business School, Columbia University}
   \city{New York}
   \state{New York}
   \country{USA}
 }

 \author{Xiao-Yang Liu Yanglet}
  \authornote{Correponding author.}
 \email{xl2427@columbia.edu }
 \affiliation{
   \institution{Columbia University}
   \city{New York}
   \state{New York}
   \country{USA}
 }

\begin{abstract}

Reinforcement learning has demonstrated great potential for performing financial tasks. However, it faces two major challenges: policy instability and sampling bottlenecks. In this paper, we revisit ensemble methods with massively parallel simulations on graphics processing units (GPUs), significantly enhancing the computational efficiency and robustness of trained models in volatile financial markets. Our approach leverages the parallel processing capability of GPUs to significantly improve the sampling speed for training ensemble models. The ensemble models combine the strengths of component agents to improve the robustness of financial decision-making strategies. We conduct experiments in both stock and cryptocurrency trading tasks to evaluate the effectiveness of our approach. Massively parallel simulation on a single GPU improves the sampling speed by up to $1,746\times$ using $2,048$ parallel environments compared to a single environment. The ensemble models have high cumulative returns and outperform some individual agents, reducing maximum drawdown by up to $4.17\%$ and improving the Sharpe ratio by up to $0.21$.

This paper describes trading tasks at ACM ICAIF FinRL Contests in 2023 and 2024.
\end{abstract}


% TODO NOT SURE WHAT TO DO HERE

\maketitle

\section{Introduction}
Advancements in reinforcement learning (RL) have led to significant breakthroughs across various domains, notably in finance, where decision-making is crucial ~\cite{hambly2021recent}. Financial reinforcement learning (FinRL) ~\cite{liu2020finrl, liu2021finrl} focuses on applying RL to perform financial tasks, such as algorithmic trading ~\cite{zhang2019deep}, portfolio management ~\cite{gu2023portfolio}, and option pricing ~\cite{vittori2021option}. Given the high computational demands of these tasks and the highly dynamic financial markets, efficient and robust solutions are essential.

However, two major challenges are encountered: policy instability and the sampling bottleneck. The challenge of policy instability significantly impacts agents' performance and reliability in RL ~\cite{rl_reliability_metrics}. Policy instability for many algorithms can come from value function approximation errors ~\cite{scott2018, sutton1999policy}. The experiments in ~\cite{henderson2018deep} show that the performance of RL policies is also sensitive to hyperparameters, unstable environments, and even random seeds. Sampling bottlenecks in RL pose significant challenges, especially for complex tasks, e.g., training robots, which require a substantial volume of samples ~\cite{rudin2022walk}. As the complexity of the task increases, the number of samples needed to train a model increases, often extending the simulation phase for robots to several days or even weeks ~\cite{rudin2022walk}. The inherent complexity and the dynamic nature of financial markets ~\cite{liu2024dynamic} further complicate the development of robust models with the added challenge of policy instability and the sampling bottleneck.

\begin{figure}[t]
  \centering
  \includegraphics[scale=0.33]{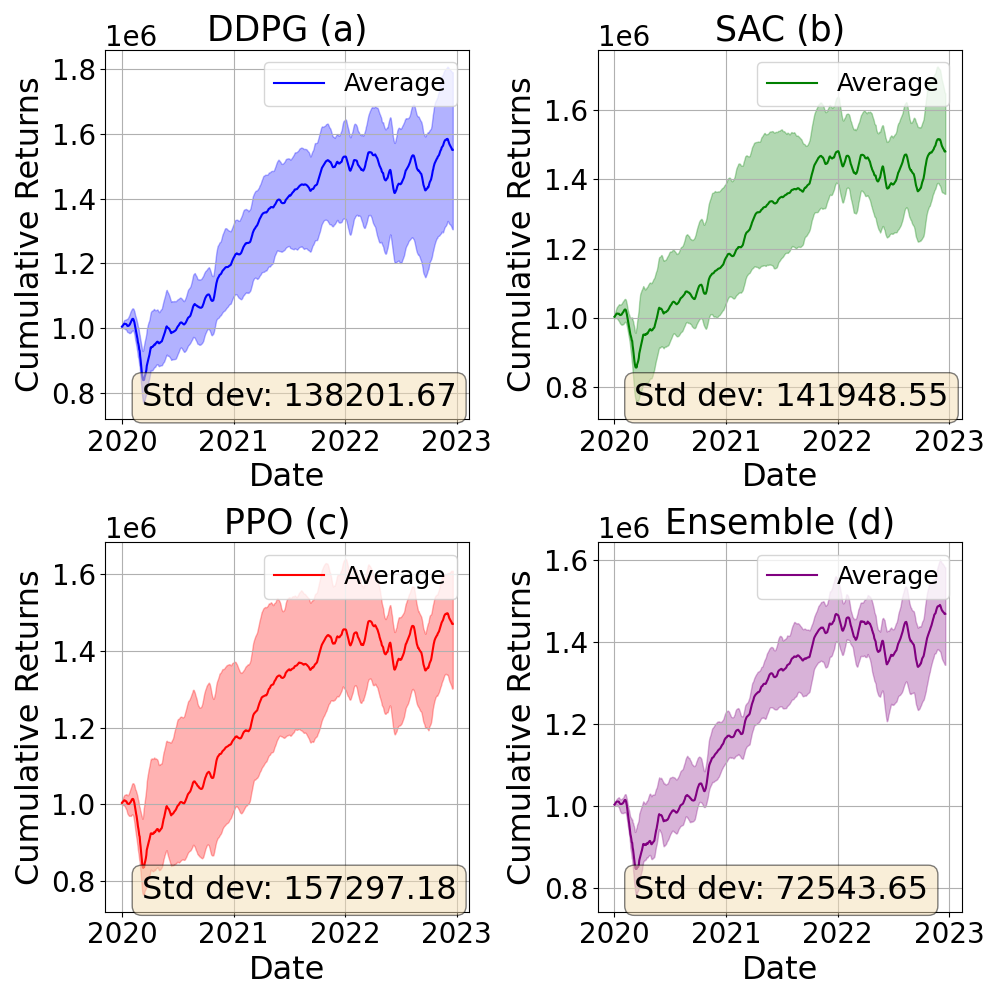}
  \vspace{-0.1in}
  \caption{Performance deviation for different RL algorithms and a simple ensemble method.}
  \vspace{-0.1in}
  \label{fig:observation_experiment}
\end{figure}

An empirical study is conducted to show policy instability. In the stock trading task, we use Proximal Policy Optimization (PPO) ~\cite{ppo}, Soft Actor-Critic (SAC) ~\cite{haarnoja2018sac}, Deep Deterministic Policy Gradient (DDPG) ~\cite{ddpg}, and an ensemble model that averages the action probabilities of three agents. The strengths and weaknesses of agents are shown in Table \ref{tab:compare_agents}. We use daily open, high, low, close, and volume (OHLCV) data for 30 Dow Jones stocks over 2020-2023. The models are trained and tested $10$ times, each time on a rolling-window basis with $30$-day training and $5$-day testing windows. Fig. \ref{fig:observation_experiment} shows cumulative returns and their standard deviation from testing. Each type of agent has a high variance in cumulative returns. In contrast, the ensemble model only has a standard deviation of returns around half that of component agents, highlighting its effectiveness in mitigating policy instability. The results support the use of ensemble methods in FinRL to enhance model robustness, though component agent training may still face the sampling bottleneck.

\begin{table}
  \caption{Strengths and weaknesses of different RL algorithms.}
  \label{tab:compare_agents}
  \begin{tabular}{lccc}
    \toprule
    Characteristics & PPO & SAC & DDPG \\
    \midrule
    Sample Efficiency & $\times $ & $\checkmark$ & $\checkmark$  \\
    Stability in Training & $\checkmark$ & $\checkmark$ & $\times$ \\
    Computational Efficiency & $\times$ & $\times$ &  $\checkmark$ \\
    Estimation Accuracy & $\checkmark$ & $\checkmark$ &  $\times$ \\
  \bottomrule
\end{tabular}
\vspace{-5.1mm}
\end{table}

In this paper, we revisit ensemble methods with massively parallel simulation, which improves the computational efficiency and robustness of models trained for volatile financial markets. First, we develop vectorized environments for massively parallel simulation in stock and cryptocurrency trading tasks. $2,048$ parallel market environments are simulated on a single GPU, and the sampling speed is improved by up to $1,746\times$ compared with a single environment. Second, we implement ensemble methods by voting on agents' actions or weighted averaging action probabilities. The ensemble models show high cumulative returns and outperform some individual agents, with a reduction of up to $4.17\%$ in maximum drawdown and an improvement of up to $0.21$ in the Sharpe ratio.
%By training multiple agents with massively parallel simulation, we are able to develop a computationally efficient and robust ensemble model, thus addressing policy instability and sampling bottleneck.

ACM ICAIF FinRL Contest 2024: website\footnote{FinRL Contest: \url{https://open-finance-lab.github.io/finrl-contest-2024.github.io/}} and Github link\footnote{FinRL Contest: \url{https://github.com/Open-Finance-Lab/FinRL_Contest_2024}}.

The remainder of this paper is organized as follows: Section 2 reviews related works. Section 3 describes the trading problem and shows the effectiveness of our approach. Section 4 develops massively parallel simulations on GPUs. Section 5 describes the ensemble method in detail. Section 6 presents performance evaluations. We conclude in Section 7 and point out our future works.

\section{Related Works}
\subsection{Financial Reinforcement Learning}
%In recent years, RL techniques have excelled in complex decision-making environments, beating human experts at challenging games and tasks such as StarCraft ~\cite{6637024}, Go and Chess ~\cite{silver2018general}. Its success has further facilitated its application in finance. 

With complex and dynamic financial markets, deep reinforcement learning (DRL) becomes more suitable for high-dimensional state and action spaces ~\cite{hambly2021recent}. DRL has been successfully applied to various financial tasks, such as algorithmic trading ~\cite{zhang2019deep}, portfolio management ~\cite{gu2023portfolio}, and option pricing ~\cite{vittori2021option}. Emerging algorithms, such as Direct Preference Optimization (DPO) ~\cite{rafailov2023direct}, which uses preferences between different outcomes as the primary feedback for learning, will also facilitate new applications of RL in finance.

\subsection{Ensemble Learning}
Ensemble methods in machine learning combine the predictions of multiple algorithms to achieve better performance than the individual components ~\cite{ensemble_ganaie}. Different ensemble methods, such as bagging ~\cite{breiman1996bagging} and boosting ~\cite{freund1996boosting}, have been developed, with many variants applied in various fields ~\cite{ensemble_ganaie}. In RL, ensemble methods can enhance overall performance by combining selected actions or action probabilities from component RL algorithms ~\cite{wiering2008ensemble}. Ensemble methods for RL are also applied in finance, such as stock trading ~\cite{yang2020deep} and cryptocurrency trading ~\cite{jing2024crypto}. In FinRL, ensemble methods can effectively enhance policy stability and agent performance in complex financial markets. 

\subsection{Simulation Environments}
RL algorithms heavily rely on simulated environments for training. For example, robot training relies on simulations because obtaining samples from the real world is costly and difficult ~\cite{ zhao2020sim}. Financial tasks also require training and validation using historical data before deployment in real-world applications ~\cite{hambly2021recent}. There are many frameworks for simulation environments in RL. OpenAI Gym ~\cite{brockman2016openaigym} is among the most popular frameworks and collections of environments. ABIDES-gym integrates the ABIDES simulator into the Gym framework and has been successfully applied to financial tasks ~\cite{amrouni2022abides}. 

Simulation environments leverage CPUs for data sampling, where the number of environments is limited to the number of CPU cores ~\cite{makoviychuk2021isaac}. Isaac Gym is developed for robot learning, which allows both the physics simulation and policy updating to occur on the GPU, thereby speeding up the training by $100\times$ to $1,000\times$ ~\cite{makoviychuk2021isaac}. For financial applications, JAX-LOB presents a GPU-accelerated limit order book (LOB) simulator designed to enable parallel processing of thousands of books on GPUs ~\cite{jaxlob}.

\section{Problem Description}

\subsection{Problem Formulation for FinRL Tasks} \label{sec:mdp}
%Financial markets are complex and dynamic, characterized by non-stationary time-series data with continually changing distributions. This makes historical datasets that are too far back less reliable for modeling and prediction~\cite{hambly2021recent}. In addition, the low signal-to-noise ratio in financial markets complicates the extraction of alpha signals, and the creation of smart beta indices ~\cite{liu2024dynamic}.
We implement stock and cryptocurrency trading tasks. Financial markets are characterized by non-stationary time-series data and low signal-to-noise ratios ~\cite{hambly2021recent, liu2024dynamic}. In stock trading, the substantial noise in data complicates the extraction of alpha signals and the creation of smart beta indices ~\cite{liu2024dynamic}. Cryptocurrency trading faces greater volatility due to drastic price fluctuations and market sentiment shifts ~\cite{jing2024crypto}. Cryptocurrency markets operate 24/7, demanding adaptable strategies in a continuous trading environment without traditional market open and close cycles. Developing robust and profitable trading strategies is crucial in these financial markets.

\begin{itemize}[leftmargin=*]
    \item \textbf{Stock trading task} involves buying and selling 30 stocks of the Dow Jones index to maximize financial returns over a specified timeframe. The trader needs to utilize daily OHLCV data to predict price movements and make trading decisions.
    \item \textbf{Cryptocurrency trading task} involves buying and selling Bitcoin (BTC) to maximize financial returns over a timeframe. The trader needs to utilize second-level LOB data to predict price movements and make trading decisions.
\end{itemize}

The two tasks, involving sequential decision-making, can be formulated as Markov Decision Processes (MDPs) ~\cite{liu2024dynamic}:
\begin{itemize}[leftmargin=*]
    \item \textbf{State} $\mathbf{s_t} = [b_t, \mathbf{p_t}, \mathbf{h_t}, \mathbf{f_t}] \in \mathbb{R}^{(I+2)K+1}$. The state at time $t$ represents the market conditions a trader might observe at time $t$. $b_t \in \mathbb{R}_+$ is the trader's account balance at time $t$. $\mathbf{p_t} \in \mathbb{R}_+^K$ is the price for each stock or cryptocurrency, where $K$ is the number of assets to trade. $\mathbf{h_t} \in \mathbb{R}_+^K$ is the holding position for each asset. $\mathbf{f_t} \in \mathbb{R}^{KI}$ is the feature vector incorporating $I$ technical indicators for each asset. Technical indicators can be common market indicators, such as Moving Average Convergence Divergence (MACD), or derived using complex methodologies to increase the signal-to-noise ratio.
    \item \textbf{Action} $\mathbf{a_t} \in \mathbb{R}^K$. The action at time $t$ is the trading action for each stock or cryptocurrency, represented as changes in positions, i.e., $\mathbf{h_{t+1}} = \mathbf{h_t} + \mathbf{a_t}$. Actions indicate changes in position rather than the position itself because limiting the range of position changes can reduce learning difficulty. An entry $a_t^i > 0, i = 1,\ldots, K$ indicates buying $a_t^i$ shares of asset $i$ at time $t$ in anticipation of price increases; $a_t^i < 0$ indicates selling $a_t^i$ shares of asset $i$ in anticipation of price declines; $a_t^i = 0$ indicates maintaining the current position.
    %The action includes buying, selling, or holding a certain quantity of each stock or cryptocurrency, i.e., $\mathbf{h_{t+1}} = \mathbf{h_t} + \mathbf{a_t}$. An entry $a_t^i > 0, i = 1,\ldots, K$, means buying $a_t^i$ shares for stock or cryptocurrency $i$ at time $t$; $a_t^i < 0$ means selling $a_t^i$ shares at time $t$; $a_t^i = 0$ means keeping the holding shares unchanged.
    \item \textbf{Reward function} $R(\mathbf{s_t}, \mathbf{a_t}, \mathbf{s_{t+1}})$. The reward, as an incentive signal, motivates the trading agent to execute action $\mathbf{a_t}$ at state $\mathbf{s_t}$. While complex rewards can be carefully designed using various financial metrics such as the Sharpe ratio and Profit and Loss (PnL) ~\cite{hambly2021recent}, we opt for a simpler reward calculation. The massive sampling approach will ensure robust learning even with a simple reward calculation. The reward is the change in total asset values, i.e., $R(\mathbf{s_t}, \mathbf{a_t}, \mathbf{s_{t+1}}) = v_{t+1} - v_{t}$, where $v_{t+1}$ and $v_{t}$ are the total asset values at time $t$ and $t+1$, respectively. We have $v_t = b_t + \mathbf{p_t}^T\mathbf{h_t}$.
    %The reward can be carefully designed, incorporating different financial metrics such as the Sharpe ratio and net Profit and Loss (PnL). For example, the reward can be defined as the change in total asset values, i.e., $R(\mathbf{s_t}, \mathbf{a_t}, \mathbf{s_{t+1}}) = v_{t+1} - v_{t}$, where $v_{t+1}$ and $v_{t}$ are the total asset values at time $t$ and $t+1$, respectively. We have $v_t = b_t + \mathbf{p_t}^T\mathbf{h_t}$.
    \item \textbf{Policy} $\pi( \cdot | \mathbf{s})$ is a probability distribution over actions at state $\mathbf{s}$. The policy assesses the current market conditions and determines the likelihood of each possible trading action.
\end{itemize}

%To obtain an optimal policy using an MDP, it is important to explore a broader area of possible transitions. To increase the exploration of agents for real-world applications, we can thus use multiple agents in ensembles.

\subsection{Training Process}
The training process can be divided into two phases: simulation and learning, following the standard \textit{Producer-Consumer} model:
\begin{itemize}[leftmargin=*]
    \item \textbf{Simulation phase}, acting as the "Producer," implements data sampling by executing the actions within the environment, resulting in new states and rewards. % state reward samples. %These samples will be stored in the replay buffer.
    \item \textbf{Replay buffer} serves as a reservoir to store samples from the simulation phase, allowing the learning phase to access these samples.
    \item \textbf{Learning phase}, acting as the "Consumer," retrieves samples from the replay buffer to update the policy. 
    %This phase analyzes past trading experiences to optimize future decision-making.
    % The synchronized policy network will be used to sample data for the next training loop. 
\end{itemize}

In the trading task, OHLCV or LOB datasets are transformed into market environments configured with realistic trading constraints, including transaction costs, slippage, turbulence threshold, and stop-loss mechanisms. During the simulation phase, the agent, acting as a trader, decides and executes the trading actions based on the current state, resulting in new states and financial outcomes quantified as rewards. The data samples, including trading actions, market states, and rewards, are stored in the replay buffer as trading experiences. In the learning phase, the agent accesses the data and learns from a wide range of trading experiences to update its policy and enhance future decision-making.

\subsection{Effectiveness and Burdens of Ensemble Methods}

\subsubsection{\textbf{Effectiveness and Costs of Ensemble Methods}}
%In financial tasks, policy instability can often be attributed to the complex and highly dynamic nature of financial markets. By increasing the number of agents, and hence increasing the number of possible trajectories explored, ensemble methods are effective in enhancing model performance and robustness ~\cite{wiering2008ensemble}. 
%The experiments in Fig. \ref{fig:observation_experiment} show the effectiveness of using ensemble methods in financial tasks.
Condorcet's theorem for voting provides a theoretical foundation for using ensemble methods in decision-making ~\cite{ensemble_ganaie}.
%Condorcet's theorem for voting indicates that if each component model in an ensemble has more than $50\%$ probability of making a correct decision, adding more models increases the probability until $100\%$ that the majority decision is correct ~\cite{ensemble_ganaie}.
In financial decision-making, consider an ensemble composed of $n$ independent traders deciding whether to buy or sell a stock, each with a probability $p$ of making the correct decision. 
% We have the probability $P$ that the majority of the traders make the correct decision:
% \begin{equation}
%     P = \sum_{k = \lceil{\frac{n}{2}}\rceil}^n \binom{n}{k}p^k (1-p)^{n-k},
% \end{equation}
% where $k$ is the number of traders with correct decisions. 
%The sum starts from $\lceil{\frac{n}{2}}\rceil$, the smallest integer greater than or equal to half of $n$ ensuring a majority.
%As the number of models $n$ increases, the probability $P$ approaches 1, provided $p > 0.5$. This result is based on the law of large numbers, as applied to a binomial distribution with the mean success rate $p > 0.5$. 
Let $X$ denote the total number of traders making the correct decision, $X \sim Binomial(n,p)$, and $\mathbb{E}(X) = np$ and $\text{Var}(X) = np(1-p)$. Using the Central Limit Theorem (CLT),
\begin{equation}
    P = \mathbb{P}\left(X > \frac{n}{2}\right) = \mathbb{P}\left(Z > \frac{\frac{n}{2}-np}{\sqrt{np(1-p)}}\right) = \mathbb{P}\left(Z > \frac{n(\frac{1}{2}-p)}{\sqrt{np(1-p)}}\right).
\end{equation}
If $p>\frac{1}{2}$, as $n\to\infty$, $\frac{n(\frac{1}{2}-p)}{\sqrt{np(1-p)}} \to -\infty$. Since $\mathbb{P}(Z > -\infty) = 1$, as $n\to\infty$, $P \to 1$. The probability that the ensemble makes the correct trading decision approaches $1$ as the number of traders with $p > 0.5$ increases. 
%In FinRL, agents act as traders and make trading decisions based on their policy. 
In FinRL, deep neural networks used for the agent's policy architecture typically have accuracies above $0.5$ ~\cite{ensemble_ganaie}, making ensemble methods appealing.

However, ensemble methods in FinRL still need to address agent diversity and extensive sampling requirements. The diversity of component agents is essential for risk mitigation by leveraging various trading strategies. Achieving high diversity requires training multiple agents across environments that simulate different market scenarios. The data-intensive nature of policy networks requires extensive sampling for effective training. Due to the sampling bottleneck, the ensemble's training time increases, making it costly and difficult to adapt quickly to volatile financial markets.

\begin{figure*}
  \centering
  \includegraphics[scale=0.38]{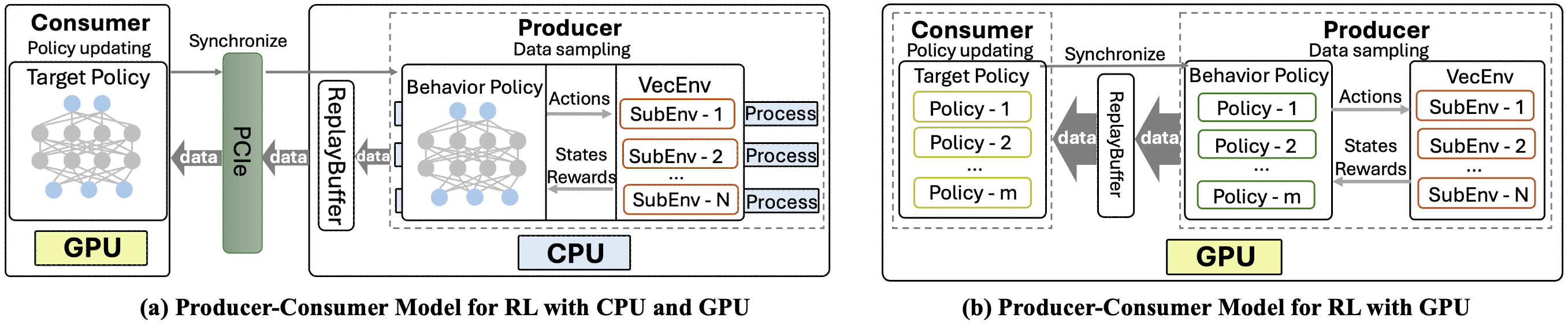}
  \vspace{-0.2in}
  \caption{Producer-Consumer model for RL.}
  \vspace{-0.1in}
  \label{fig:producer_consumer}
\end{figure*}

\subsubsection{\textbf{Challenge of Extensive Sampling}}
The goal is to learn a policy $\pi_{\theta}$ with parameter $\theta$ that maximizes the expected return:
\begin{equation}
    J(\theta) = \int_{\tau} P(\tau|\pi)R(\tau) = \mathbb{E}_{\tau \sim \pi_{\theta}}[R(\tau)],
\end{equation}
where $\tau$ is a trajectory, $R(\tau)$ is the (discounted) cumulative return along the trajectory $\tau$. The gradient of $J(\theta)$ with respect to $\theta$ is ~\cite{peters2010gradient}:
\begin{equation}
\label{eq:gradient_J}
    \nabla J(\theta) = \mathbb{E}_{\tau \sim \pi_{\theta}}\left[ \sum_{t=1}^{T}  R(\tau) \nabla_{\theta} \log \pi_{\theta}(a_t|s_t) \right].
\end{equation}

In our FinRL tasks, a trajectory $\tau$ is a sequence of trading actions and states observed over a period. Financial outcomes are quantified as rewards along $\tau$ to compute $R(\tau)$. The trading strategy is governed by $\pi_{\theta}$ to maximize $J(\theta)$. When dealing with complex and noisy financial datasets, achieving a low-variance gradient estimation $\nabla J(\theta)$ is crucial for stable and reliable policy updates, reducing the risk of suboptimal trading strategies. Due to the sensitivity of financial rewards to small changes in actions, extensive sampling is necessary to reduce variance in gradient estimation $\nabla J(\theta)$.

%Small changes in action can lead to vastly different rewards, resulting in extremely high variance; extensive sampling is required to reduce variance in gradient estimation $\nabla J(\theta)$. 

\section{Massively Parallel Simulation} \label{sec:massively}
\subsection{Simulation Phase for Gradient Estimate}
To estimate $\nabla J(\theta)$ in (\ref{eq:gradient_J}), we can use the Monte Carlo method \cite{monte_carlo}:
\begin{equation}
    \label{eq:gradient_esitmate}
    \nabla J(\theta) = \frac{1}{N} \sum_{i=1}^{N} \sum_{t=1}^{T} R(\tau^{(i)}) ~\nabla_{\theta} \log \pi_{\theta}(a_t^{(i)}|s_t^{(i)}), 
\end{equation}
Where $N$ trajectories are used. The Law of Large Numbers guarantees that as the sample size $N$ increases, the estimation of $ \nabla J(\theta)$ will converge to its expected value. According to CLT, increasing $N$ leads to a reduction in the variance of the estimate.

\subsection{Massively Parallel Market Environments}
\subsubsection{\textbf{Parallelsim of Simulation Phase}}
As shown in (\ref{eq:gradient_esitmate}), a large number $N$ of trajectories sampled during the simulation phase is required to reduce the variance of $\nabla J(\theta)$. 
%There exists a high degree of parallelism in $N$. 
In (\ref{eq:gradient_esitmate}), each $i$ in the outer sum from $1$ to $N$ corresponds to a separate trajectory $\tau^{(i)}$, which can be considered as a complete and independent simulation of the policy $\pi_\theta$ in the environment. Therefore, each trajectory $\tau^{(i)}$ can be simulated in parallel, allowing for a high degree of parallelism. In addition, the degree of parallelism scales with the processing capability of computational resources, which is crucial for implementing massively parallel simulations.

In FinRL, parallel simulation involves executing the trading strategy in multiple market scenarios simultaneously. The parallelism accelerates the simulation phase, allowing for more rapid updates and iterations of the policy $\pi_{\theta}$. Therefore, the trading strategy governed by $\pi_{\theta}$ can be quickly updated and adapted to changing market conditions. 

%In addition, parallelism scales with the availability of computational resources. With more parallel processing capability, more trajectories can be computed simultaneously. Such scalability is crucial for implementing massively parallel simulations, effectively addressing the sampling bottleneck.

\subsubsection{\textbf{Vectorized Market Environments}}
We develop vectorized market environments for massively parallel simulation. 

\textbf{Parallel sub-environments}. As shown in Fig. \ref{fig:producer_consumer}, a vectorized environment (VecEnv) manages parallel sub-environments (SubEnv). Each SubEnv simulates different market scenarios using diverse OHLCV or LOB datasets, maintaining its own balances, prices, holding positions, technical factors, and market constraints. As the demand for data sampling grows, more SubEnvs can be added, enhancing parallelism and computational efficiency.

\textbf{Building environments}. We perform consistent operations across all SubEnvs for data sampling:
%We define the functions \texttt{reset()}, \texttt{step()}, and \texttt{reward()} in the vectorized environment as follows:
\begin{itemize}[leftmargin=*]
    \item \texttt{reset}: $s_t \rightarrow s_0$, resets the environment to its initial state. It resets to the initial market conditions, with all variables, such as asset prices and balances, set to their starting values.
    \item \texttt{step}: ($s_t, a_t) \rightarrow s_{t+1}$, takes action $a_t$ and updates $s_t$ to $s_{t+1}$. It executes the trade in the market, resulting in a new market state.
    \item \texttt{reward}: $(s_t, a_t, s_{t+1}) \rightarrow r_t$, computes the reward. As defined in Section \ref{sec:mdp}, the reward measures the financial performance of the trading action taken in the current market. 
\end{itemize}

%\textbf{Environment scalability}. We have a uniform structure of sub-environments and consistent \texttt{reset}, \texttt{step}, and \texttt{reward} implementations for simulation. This architecture of vectorized environments supports scalability. As the demand for data sampling grows, more sub-environments can be added, thereby increasing parallelism without changing the fundamental VecEnv architecture.

\subsection{Mapping onto GPUs}
%The implementation of massively parallel simulation on the GPU is a cornerstone of our approach to accelerating data sampling in FinRL. 

\subsubsection{\textbf{Parallel Simulations on GPUs}}
Modern GPUs have high parallel processing capabilities, making them well-suited for massively parallel simulation across many GPU cores. The \texttt{vmap} function of PyTorch vectorizes the \texttt{step} and \texttt{reward} functions, enabling them to operate simultaneously in thousands of parallel SubEnvs. An operation on multiple data points from SubEnvs will be efficiently dispatched across available GPU cores. In FinRL, for example, when calculating the financial performance of trading actions in all SubEnvs, the \texttt{reward} function, vectorized by \texttt{vmap}, executes computations on $(s_t, a_t, s_{t+1})$ from all SubEnvs. This computation will be dispatched to available GPU cores, with each core responsible for calculating its assigned data.

\begin{figure*}[t]
  \centering
  \includegraphics[scale=0.38]{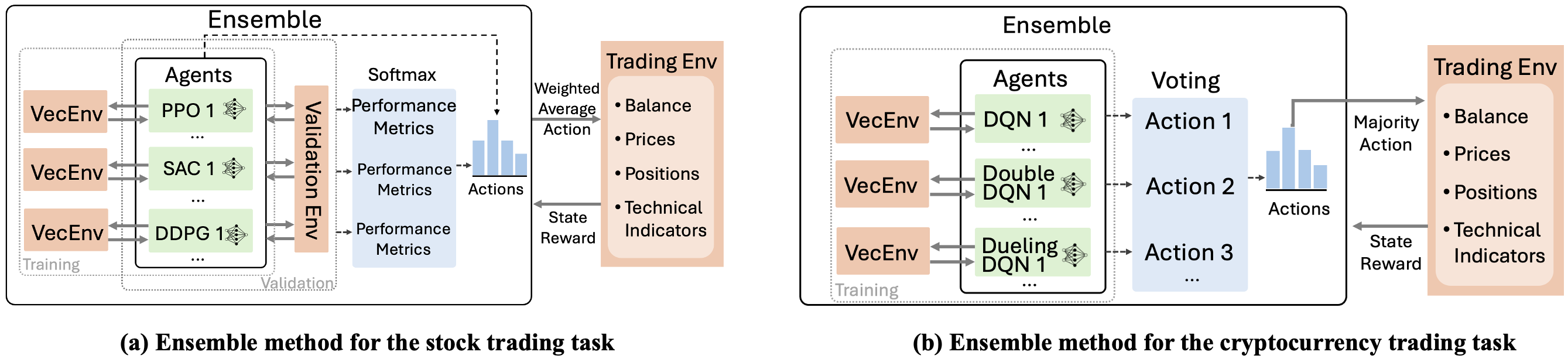}
  \vspace{-0.25in}
  \caption{Ensemble methods.}
  \vspace{-0.1in}
  \label{fig:ensemble_method}
\end{figure*}

\subsubsection{\textbf{Storing Data Samples in GPU Tensors}}
The data samples are organized into tensors and stored in GPU memory. The tensors for states, actions, and rewards have the shape $T \times N \times D$:
\begin{itemize}[leftmargin=*]
    \item $\mathbf{T}$ is the number of steps in a trajectory.
    \item $\mathbf{N}$ is the number of parallel SubEnvs in a VecEnv.
    \item $\mathbf{D}$ is the dimension as specified in Section \ref{sec:mdp}.
    %we have $D_s = (I+2)K+1$ for state, $D_a = K$ for action, and $D_r = 1$ for reward, where $K$ is the number of stocks or cryptocurrencies to trade and each stock or cryptocurrency has $I$ technical indicators.
\end{itemize}
Therefore, we have tensors for states ($\mathbf{s} \in \mathbb{R}^{D_s}$), actions ($ \mathbf{a} \in \mathbb{R}^{D_a}$), and rewards ($r \in \mathbb{R}^{D_r}$) like the following:

\[
\begin{bmatrix}
    %\mathbf{s}_0^1 & \mathbf{s}_0^2 & \cdots & \mathbf{s}_0^N \\
    \mathbf{s}_1^1 & \mathbf{s}_1^2 & \cdots & \mathbf{s}_1^N \\
    \mathbf{s}_2^1 & \mathbf{s}_2^2 & \cdots & \mathbf{s}_2^N \\
    \vdots & \vdots & \ddots & \vdots \\
    \mathbf{s}_T^1 & \mathbf{s}_T^2 & \cdots & \mathbf{s}_T^N \\
\end{bmatrix},
\begin{bmatrix}
    %\mathbf{a}_0^1 & \mathbf{a}_0^2 & \cdots & \mathbf{a}_0^N \\
    \mathbf{a}_1^1 & \mathbf{a}_1^2 & \cdots & \mathbf{a}_1^N \\
    \mathbf{a}_2^1 & \mathbf{a}_2^2 & \cdots & \mathbf{a}_2^N \\
    \vdots & \vdots & \ddots & \vdots \\
    \mathbf{a}_T^1 & \mathbf{a}_T^2 & \cdots & \mathbf{a}_T^N \\
\end{bmatrix},
\begin{bmatrix}
    %r_0^1 & r_0^2 & \cdots & r_0^N \\
    r_1^1 & r_1^2 & \cdots & r_1^N \\
    r_2^1 & r_2^2 & \cdots & r_2^N \\
    \vdots & \vdots & \ddots & \vdots \\
    r_T^1 & r_T^2 & \cdots & r_T^N \\
\end{bmatrix}
\]
A trajectory sampled from the $i$th environment is 
%$(\mathbf{s}_0^i, \mathbf{a}_0^i, r_0^i, \mathbf{s}_1^i, \mathbf{a}_1^i, r_1^i, \\ \ldots, \mathbf{s}_T^i)$.
$(\mathbf{s}_0^i, \mathbf{a}_1^i, \mathbf{s}_1^i, \mathbf{a}_2^i, \ldots, \\ \mathbf{s}_T^i)$, where $\mathbf{s}_0^i$ is the initial state. The cumulative return along this trajectory is $R(\tau^{(i)}) = \sum_{t=1}^T \gamma^t r_t^{(i)}$, where $\gamma$ is the discount factor.

Storing data samples in tensors in GPU memory avoids the costly CPU-GPU communication.
%When data samples are stored in tensors in GPU memory, all data transferring and processing occur on GPU, avoiding the costly communication between CPU and GPU. 
Fig. \ref{fig:producer_consumer} (a) shows a traditional training process with both the CPU and GPU. The simulation phase, managed by the "Producer," performs data sampling on the CPU. The data samples are stored in a replay buffer on the CPU. The learning phase, managed by the "Consumer" and typically run on GPUs, fetches the data on the CPU for policy updating. PCIe allows for communication between the CPU and GPU but has limited bandwidth, making frequent large data transfers a significant bottleneck. This problem can be solved by storing and processing data samples in tensors on the GPU. As shown in Fig. \ref{fig:producer_consumer} (b), the end-to-end GPU-accelerated agent training framework ~\cite{makoviychuk2021isaac} is used, where both the simulation and learning phases are conducted on the GPU. It avoids the bandwidth limitations of PCIe, thereby reducing the latency of GPU-CPU communication and addressing the sampling bottleneck. 

\section{Ensemble Learning}

\subsection{Agent Diversity} \label{sec:agent_diversity}
%As shown in Fig. \ref{fig:ensemble_method}, for different tasks, we have various types of agents, each with multiple instances, to ensure a large enough number of agents. 
%We use various approaches to ensure agent diversity, which is crucial to enhancing the ensemble's performance.

\textbf{Using KL divergence in objective functions}. To enforce diversity among the component agents, we introduce a Kullback-Leibler (KL) divergence term into the agent's training loss function. The KL divergence measures how one probability distribution diverges from another ~\cite{perez2008kldivergence}. In the ensemble, the KL divergence term penalizes similarities in policies between different agents, encouraging them to adopt various trading strategies. The new training loss function for a component agent is as follows:
\begin{equation}
    L_{\text{new}}(\theta_i)=L_{\text{original}}(\theta_i) - \lambda \sum_{j \neq i} \text{KL}(\pi_{\theta_j}||\pi_{\theta_i}),
\end{equation}
where $\theta_i$ are the policy parameters for agent $i$, $L(\theta_i)$ is the training loss, $\text{KL}(\pi_{\theta_j}||\pi_{\theta_i})$ is the KL divergence between agent $i$'s and agent $j$'s policies, and $\lambda$ is a regularization constant.

\textbf{Using various datasets}. The financial datasets used for training component agents are varied. For each stock or cryptocurrency, a random percentage change ranging from $-1\%$ to $1\%$ is generated and applied to its prices, which shifts the price scale while preserving the original price trends.
%It helps prevent models from overfitting. 
%We also use different periods of data to train the component agents. 
Agents are also trained on different stocks from the test set for the stock trading task. It enables agents to learn various strategies for a broader range of stocks rather than reacting to a limited number of stocks.

\subsection{Ensemble Methods for FinRL Tasks}
% As shown in Fig. \ref{fig:ensemble_method}, we use different agents and ensemble methods for stock and cryptocurrency trading tasks.

\subsubsection{\textbf{Stock Trading Task}} \label{sec:stock_ensemble}
As shown in Fig. \ref{fig:ensemble_method} (a), the ensemble includes PPO, SAC, and DDPG agents, with their strengths shown in Table \ref{tab:compare_agents}. The ensemble's final trading action is determined by weighted averaging over the agents' action probabilities. The process is as follows:
\begin{itemize}[leftmargin=*]
    \item \textbf{Traning}. Agents are trained independently with a VecEnv on a 30-day training rolling window, using massively parallel simulation in Section \ref{sec:massively} and agent diversity methods in Section \ref{sec:agent_diversity}. 
    % We use different strategies to ensure agent diversity as described in \ref{sec:agent_diversity}.
    \item \textbf{Validation}. After training, agents are validated on a 5-day rolling window. Sharpe ratios are calculated to evaluate their ability to balance returns with associated risks.
    \begin{equation} \label{sharpe_ratio}
        \text{Sharpe Ratio} = \frac{\bar{r}_p - r_f}{\sigma_p},
    \end{equation}
    where $\bar{r}_p$ is the portfolio return, $r_f$ is a chosen risk-free rate, and $\sigma_p$ is the standard deviation of the portfolio return.
    \item \textbf{Weights calculation}. Agents with very low Sharpe ratios are discarded. Weights for the remaining agents are calculated using a softmax function applied to their Sharpe ratios.
    \item \textbf{Trading}. The ensemble acts based on a weighted average of agent action probabilities during a 5-day trading window.
\end{itemize}
%After each loop, the training, validation, and trading windows are rolled forward. 
This rolling window approach ensures that the ensemble method remains adaptive to the continuously changing market. 

%The stock trading task is designed as a portfolio trading task, where the model predicts the number of shares to buy, sell, and hold for any stock at each timestep. As seen in Fig. \ref{fig:ensemble_method} we use PPO, SAC ~\cite{ haarnoja2018sac}, and DDPG \cite{ddpg} agents. These models were trained independently on the in-sample data to develop distinct trading strategies using a rolling window where the ensemble agents are trained on $30$ days of data, then evaluated on $5$ out of sample days and then tested on $5$ more out of sample days on a repeating basis. Post-training, agents were validated based on their ability to balance returns with associated risks. Weights for each agent’s outputs in the ensemble were assigned using a softmax function applied to the Sharpe ratios obtained during this validation phase. 

\subsubsection{\textbf{Cryptocurrency Trading Task}} \label{sec:crypto_ensemble}
For cryptocurrency trading at a relatively high frequency, market movements can be modeled as discrete events, which require a discrete action space. As shown in Fig. \ref{fig:ensemble_method} (b), DQN ~\cite{Mnih2015}, Double DQN ~\cite{doubledqn}, and Dueling DQN ~\cite{duelingdqn} are used to handle this discrete action space. In addition, the dataset for a single cryptocurrency is relatively small. DQN and its variants, with fewer parameters and simpler architectures, can be trained faster to avoid overfitting. Moreover, trading at a high frequency requires fast responses, and DQN agents can offer lower latency in decision-making compared to more complex models. The ensemble model uses majority voting to combine the actions of component agents. Majority voting ensures the chosen action reflects consensus among agents, mitigating biases from any single agent's actions ~\cite{ensemble_ganaie}. The process is as follows:
\begin{itemize}[leftmargin=*]
    \item \textbf{Training}. Each component agent is independently trained with a VecEnv, using the massively parallel simulation in Section \ref{sec:massively} and the agent diversity methods in Section \ref{sec:agent_diversity}. 
    \item \textbf{Action ensemble and trading}. During the trading phase, each agent processes the same market state and determines an action based on its policy. The majority action is selected as the final ensemble action.
\end{itemize}

\section{Performance Evaluations}

\begin{figure*}
  \centering
  \includegraphics[scale=0.40]{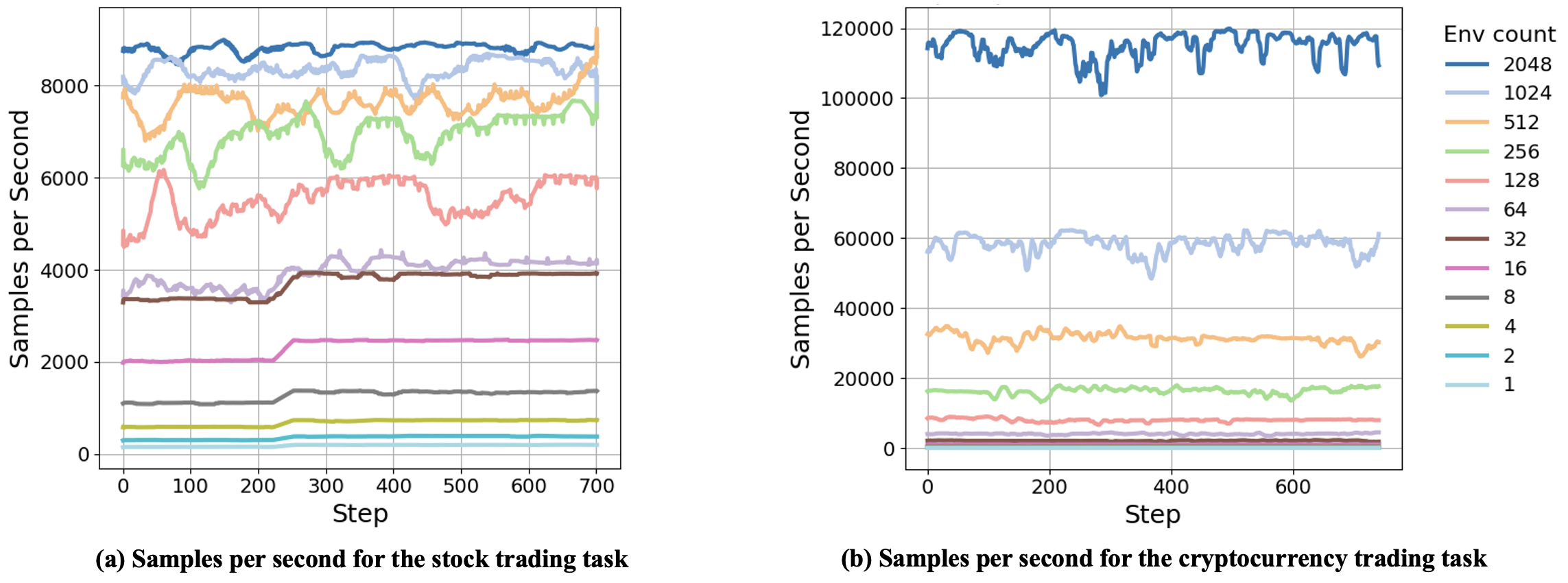}
  \vspace{-0.15in}
  \caption{Samples per second for the stock trading task and the cryptocurrency trading task. NVIDIA A100 GPU is used.}
  \vspace{-0.1in}
  \label{fig:samples_stock}
\end{figure*}

\subsection{Experiement Settings}

All experiments were conducted using one NVIDIA A100 GPU.

\textbf{Stock data}: We use historical daily OHLCV data for all 30 stocks in the Dow Jones index from 01/01/2021 to 12/01/2023. OHLCV data is a rich source for learning financial market behaviors and trends. We use Moving Average Convergence Divergence (MACD), Bollinger Bands, Relative Strength Index (RSI), Commodity Channel Index (CCI), Directional Movement Index (DX), and Simple Moving Average (SMA) as technical indicators. These indicators enrich the data with more insights into market behaviors and trends.

\textbf{Cryptocurrency data}: The dataset comprises second-level LOB data for BTC from 04/07/2021 11:32:42 to 04/19/2021 09:54:22. This dataset provides a detailed view of buying and selling activities in the BTC market, enabling analysis of market dynamics. Adaptations from 101 formulaic alphas ~\cite{alpha101} are calculated based on LOB data to extract insights into market behaviors, such as momentum, mean-reversion, and market anomalies. A recurrent neural network (RNN) further processes the 101 alphas into $8$ technical indicators. It reduces the complexity of the input data and enhances the ability to predict market trends, thus improving generalization and avoiding overfitting.

%To enhance generalization and avoid overfitting, a recurrent neural network (RNN) further processes 101 alphas into $8$ robust technical indicators, reducing the complexity of the input data and enhancing the model’s ability to predict market trends effectively.
%To avoid overfitting, a recurrent neural network (RNN) further processed these alphas, transforming them into $8$ strong technical indicators to effectively predict market trends. 
%After preprocessing the data, the dataset was split into two distinct periods: (1) the in-sample period used for training, and (2) the out-of-sample period used for trading.

\textbf{Agents for stock trading task}. We use PPO, SAC, and DDPG agents. The policy network for each agent consists of a feed-forward network with two hidden layers, having 64 units and 32 units, respectively. We set a learning rate of $3\cdot10^{-4}$ and a batch size of $64$. The agents are trained, validated, and tested on a rolling-window basis with 30-day training, 5-day validation, and 5-day testing windows. There are no leaks of future information.

\textbf{Agents for cryptocurrency trading task}.
%the RNN uses parallel LSTM and GRU layers and combines their outputs when processing 101 alphas. 
We use DQN, Double DQN, and Dueling DQN agents. The policy network for each agent consists of a feed-forward neural network with three 128-unit hidden layers. We set an exploration rate of $0.005$, a learning rate of $2\cdot 10^{-6}$, and a batch size of $512$. The RNN is trained on data from 04/07/2021 11:32:42 to 04/17/2021 00:38:02 without future information leaks. The agents are trained on the in-sample data from 04/17 00:38:03 to 04/19 09:09:21 and tested on the out-of-sample data from 04/19 09:09:22 to 04/19 09:54:22.

\textbf{Performance metrics}: We evaluate the performance using the following metrics:
\begin{itemize}[leftmargin=*]
    \item \textbf{Cumulative return} is the total return generated by the trading strategy over a trading period.
    \item \textbf{Annual return} is the geometric average amount of money earned by the agent each year over a given time period.
    \item \textbf{Annual volatility} is the annualized standard deviation of daily returns.
    \item \textbf{Sharpe ratio} measures the excess return per unit of volatility, as defined in (\ref{sharpe_ratio}).
    \item \textbf{Maximum drawdown} measures the largest single drop in the portfolio value from peak to trough.
    \item \textbf{Return over maximum drawdown (RoMaD)} is calculated as the cumulative return divided by the maximum drawdown.
    \item \textbf{Sortino ratio} is calculated as the excess return divided by the downside deviation.
    \item \textbf{Calmar ratio} is calculated as the annualized excess return divided by the maximum drawdown.
    \item \textbf{Omega ratio} compares the probability of achieving returns above a threshold to the probability of falling below it.
\end{itemize}
Cumulative and annual returns focus on the returns generated by a trading strategy. Annual volatility and maximum drawdown evaluate the risk associated with the trading strategy. The Sharpe ratio, RoMad, Sortino ratio, Calmar ratio, and Omega ratio evaluate risk-adjusted returns in different ways.

\textbf{Baselines}: We use different market indexes and trading strategies as baselines:
\begin{itemize}[leftmargin=*]
    \item \textbf{DJIA}: Dow Jones Industrial Average index for the stock market.
    %\item \textbf{BPI}: CoinDesk Bitcoin Price Index for the BTC market.
    \item \textbf{BTC price}: We use the Bitcoin price from CoinBase.
    \item \textbf{Min-variance strategy} for stock trading.
    \item \textbf{Fixed-time exit strategy} for cryptocurrency trading, entering the market when the RNN prediction for the next $225$ seconds exceeds a threshold and exiting the market after $225$ seconds.
\end{itemize}
% \begin{itemize}
%     \item Market index:
%     \begin{itemize}
%         \item Dow Jones Industrial Average (DJIA)
%         \item CoinDesk Bitcoin Price Index (BPI)
%     \end{itemize}
%     \item Baseline methods:
%     \begin{itemize}
%         \item Mean-Variance approach for stock trading.
%         \item Fixed-time exit strategy for cryptocurrency trading, entering the market when RNN predictions exceed the threshold and exiting the market after a fixed time.
%     \end{itemize}    
% \end{itemize}

\subsection{Sampling Speed}

We use a PPO agent to perform the stock trading task and a DQN agent to perform the cryptocurrency trading task. We vary the numbers of parallel environments from $1$, $2$, $4$, $\ldots$, and $2,048$. The sampling speed is measured in samples per second and plotted against training steps on the Y-axis.

%We measure the sampling speed in the simulation by calculating the samples per second at each step of training. The samples per second measure the throughput efficiency in massively parallel simulation.
%The samples per second account for the training time tradeoff of massively parallelized environments, making it a better judge of performance gain. 
%We use the PPO agent for this experiment, which updates the policy based on data sampled from the current policy. Higher sampling speeds benefit these algorithms by ensuring data remains relevant and timely for effective updates. An NVIDIA A100 GPU is used.
%Analyzing this for many steps allows us to gauge how computationally stable the massive parallelization is over time and enables us to explore compute performance tradeoffs between varying levels of parallelization. 

As shown in Fig. \ref{fig:samples_stock} (a), in the stock trading task, the simulation with $2,048$ parallel environments has an average sampling speed of $8,813.81$ samples per second. Compared to a single environment with $184.63$ samples per second, the sampling speed is improved by $47.73\times$. In the cryptocurrency trading task, as shown in Fig. \ref{fig:samples_stock} (b), the simulation with $2,048$ parallel environments achieves around $114,885.98$ samples per second versus $65.79$ samples per second in a single environment. The sampling speed is improved by $1,746\times$. This greater improvement in sampling speed compared to stock trading is due to the simpler environment and calculations for single-asset trading. The results show that massively parallel simulation greatly improves the sampling speed for FinRL tasks.

For the stock trading task, we use daily open, high, low, close, and volume (OHLCV) data for the 30 stocks in the Dow Jones index, from January 1, 2020, to January 1, 2023. For the cryptocurrency trading task, we use the limit order book data with a frequency of seconds for Bitcoin, from April 7, 2021, at 11:32:42, to April 19, 2021, at 09:54:22. Since we are not testing or validating we use data exclusively for simulation and sampling rate benchmarking.

\subsection{Stock Trading Task}
\label{experiment_stock}

\begin{table*}[]
  \caption{Stock trading task performance. Models are trained, validated, and tested on a rolling window basis on OHLCV datasets for 30 Dow Jones stocks. Ensemble models use weighted averages on agent action probabilities.}
  \label{tab:ensemble_performance_stock}
  \vspace{-0.1in}
  \begin{tabular}{lcccccccc}
    \toprule
    Model & Ensemble-1 & Ensemble-2 & Ensemble-3 & PPO & SAC & DDPG & Min-Variance & DJIA \\
    \midrule
    Cumulative Return & 62.60\% & 58.77\% & 46.89\% & \textbf{63.37}\% & 50.62\%  & 63.19\% & 13.9\%\% & 18.95\% \\
    Annual Return & 18.22\% & 17.25\% & 14.15\% & \textbf{18.41} \% & 15.14\% & 18.36\% & 7.34\% & 6.15\% \\
    Annual Volatility & 11.76\% & 12.61\% & 12.70\% & \textbf{11.35}\% & 11.67\% & 11.93\% & 18.16\% & 15.14\%\\
    Sharpe Ratio & 1.48 & 1.33 & 1.11 & \textbf{1.55} & 1.27 & 1.47 & 0.48 & 0.47 \\
    Sortino Ratio & 2.34 & 2.14 & 1.74 & \textbf{2.44} & 2.05 & 2.37 & 0.73 & 0.67\\
    Max Drawdown & \textbf{-8.98\%} & -11.27\% & -12.27\%  & -9.96\%  & -12.02\%  & -13.15\% & -14.9\% & -21.94\% \\
    RoMaD & \textbf{6.97} & 5.22 & 3.82 & 6.36 & 4.21 & 4.81 & 1.10 & 0.86 \\
    Calmar Ratio & \textbf{2.03} & 1.53 & 1.15 & 1.85 & 1.26 & 1.40 & 0.49 & 0.28 \\
    Omega Ratio & 1.31 & 1.28 & 1.23 & \textbf{1.33} & 1.27 & 1.32 & 1.09 & 1.08 \\
  \bottomrule
\end{tabular}
\end{table*}

\begin{table*}[t]
  \caption{Cryptocurrency trading task performance. The second-level LOB data for Bitcoin is split into out-of-sample data for training and in-sample data for testing. Ensemble models use majority voting on agent actions.}
  \label{tab:ensemble_performance_crypto}
  \vspace{-0.1in}
  \begin{tabular}{lcccccccc}  
    \toprule
    Model           & Ensemble-1            & Ensemble-2            & Ensemble-3        & DQN     & Double DQN   & Dueling DQN & Fixed-Time Exit & BTC Price \\
    \midrule
    Cumulative Return   & 0.66\%            & 0.66\%                & 0.66\%            & 0.34\%    & 0.48\%    & 0.48\%    & -0.1\%    & \textbf{0.74\%} \\
    Sharpe Ratio        & \textbf{0.28}      & \textbf{0.28}       & \textbf{0.28}      & 0.15      & 0.21     & 0.21      & -0.03     & 0.20 \\
    Maximum Drawdown        & \textbf{-0.73\%}   & \textbf{-0.73\%}   & \textbf{-0.73\%}   & -0.93\%   & -0.98\%   & -0.98\%   & -1.00\%   & -1.3\%  \\
    RoMaD               & \textbf{0.90}     & \textbf{0.90}      & \textbf{0.90}      & 0.37      & 0.49      & 0.49      & 0.10      & 0.59 \\
    Sortino Ratio       & \textbf{0.39}      & \textbf{0.39}      & \textbf{0.39}      & 0.20      & 0.29     & 0.29      & -0.04     & 0.28 \\
    Omega Ratio         & \textbf{1.08}      & \textbf{1.08}      & \textbf{1.08}       & 1.04      & 1.05      & 1.05      & 0.99      & 1.05\\
    %Win Rate            & \textbf{61.86\%}   & \textbf{61.86\%}   & \textbf{61.86\%}   & 56.70\%   & 61.79\%   & 61.79\%   & 4.00\%            & - \\
    %Loss Rate           & 38.14\%           & 38.14\%               & 38.14\%            & 43.30\%          & 38.21\%  & 38.21\%   & \textbf{5.10\%}    & - \\
    Win/Loss Ratio & \textbf{1.622}  & \textbf{1.622}  & \textbf{1.622} & 1.309 & 1.617 & 1.617 & 0.5 & - \\
  \bottomrule
\end{tabular}
\vspace{-0.1in}
\end{table*}

\begin{figure*}[t]
  \centering
  \includegraphics[scale=0.38]{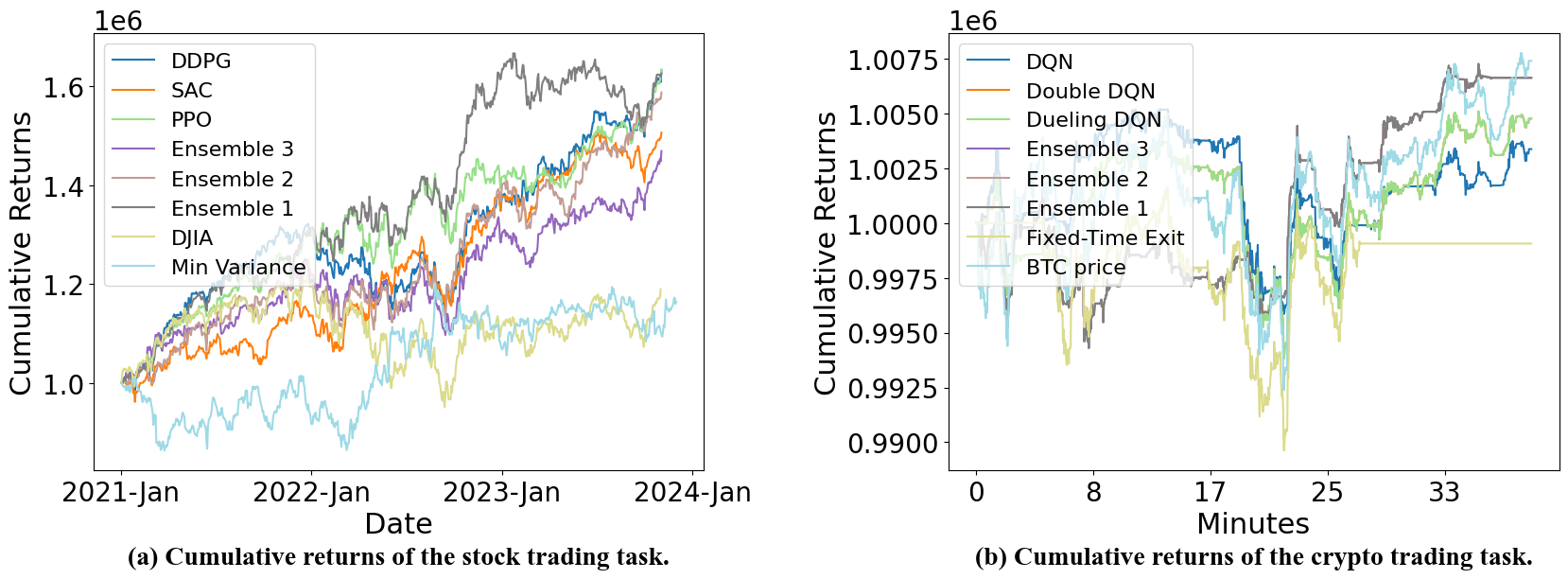}
  \vspace{-0.1in}
  \caption{Cumulative returns of different strategies for the stock trading task and cryptocurrency trading task.}
  \vspace{-0.1in}
  \label{fig:stock_returns}
\end{figure*}

We performed the stock trading task for 30 stocks in the Dow Jones index by using three ensemble models, and individual PPO, SAC, and DDPG agents. We demonstrate that ensemble agents trained in massively parallel environments can explore large spaces and find optimal strategies quickly. 

\textbf{Ensemble methods}. The first method (Ensemble 1) consists of $1$ PPO, $1$ SAC, and $1$ DDPG agents; the second method (Ensemble 2) consists of $5$ PPO, $5$ SAC, and $5$ DDPG agents; the third method (Ensemble 3) consists of $10$ agents for each type. As in Section \ref{sec:stock_ensemble}, all three ensemble models use the weighted average approach to combine component agent action probabilities. All ensemble models and individual agents are trained, validated, and tested on \textbf{a rolling-window basis} with 30-day training, 5-day validation, and 5-day testing windows.

Ensemble 1 consists of $1$ PPO and $1$ SAC agents, Ensemble 2 consists of $5$ PPO and $5$ SAC agents, and Ensemble 3 consists of $10$ PPO and $10$ SAC agents. The ensemble method is described in Section \ref{sec:stock_ensemble}. The three ensemble models and individual PPO and SAC agents are trained, validated, and tested on a rolling-window basis, with 30-day training windows, 5-day validation windows, and 5-day testing windows.  

\textbf{Results}.  As seen in Table \ref{tab:ensemble_performance_stock}, the PPO agent achieves the highest cumulative returns of $63.37\%$, Sharpe ratio of $1.55$, and Sortino ratio of $2.44$, showing an ability to maintain high returns with controlled volatility and downside risk. Although DDPG’s cumulative returns are comparable to PPO's, its higher maximum drawdown of $-13.15\%$ signals a greater risk of large value drops, which is a concern for risk management. SAC has a lower maximum drawdown than DDPG but underperforms in other metrics. All individual agents significantly outperform two traditional baselines across all metrics. The ensemble models also maintain profitability and risk management advantages over the baselines. Ensemble 1 has a high cumulative return of $62.60\%$, and as shown in Fig. \ref{fig:stock_returns} (a), it shows superior performance from Sep 2022 to Oct 2023. Ensemble 1 also achieves the smallest maximum drawdown and a higher Sharpe ratio than SAC and DDPG. Ensemble 1 and 2 have high RoMaD and Calmar ratios, showing an ability to quickly recover from peak-to-trough losses and a potential for steady growth in market adversities.

DDPG’s cumulative returns are close to PPO’s, but it has a higher max drawdown at $-13.15\%$, indicating a potential for significant value drops. SAC has a lower max drawdown than DDPG but underperforms in other metrics. All individual agents greatly outperform the mean-variance approach and DJIA in all metrics. The ensemble models effectively combine the strengths of component agents, maintaining profitability and risk management advantages. Ensemble 1 has a high cumulative return of $62.60\%$, close to the performance of PPO and DDPG. As shown in Fig. \ref{fig:stock_returns}, Ensemble 1 shows superior performance during steps 450 to 700. Additionally, Ensemble 1 and 2 minimize loss effectively during downturns, underscored by the high RoMaD and Calmar ratios.

As seen in Table \ref{tab:ensemble_performance_stock}, the ensemble method has the highest cumulative return, Sharpe Ratio, and risk-adjusted returns. We weighted agent actions according to the Sharpe Ratio, which explains the comparatively high Sharpe Ratio and risk-adjusted returns compared to the other agents. The high cumulative returns indicate that the greater number of strategies cover a broader space of market conditions as opposed to the single agent configurations which seem to explore a less diverse space of the environment. %Interestingly the single policies have slightly lower maximum drawdown than the ensemble which may be due to a noi

\subsection{Cryptocurrency Trading Task}

The cryptocurrency trading task for Bitcoin (BTC) is performed using three ensemble models, and individual DQN, Double DQN, and Dueling DQN agents.

\textbf{\textbf{Ensemble methods}} The first method (Ensemble 1) consists of $1$ DQN, $1$ Double DQN, and $1$ Dueling DQN agents; the second method (Ensemble 2) consists of $3$ DQN, $3$ Double DQN, and $3$ Dueling DQN agents; the third method (Ensemble 3) consists of $10$ agents for each type. As in Section \ref{sec:crypto_ensemble}, three ensemble models use a majority voting approach to aggregate the agents' actions. 
All ensemble models and individual agents are trained on the in-sample data and tested on the out-of-sample data. 

We evaluate their performance using cumulative returns, Sharpe ratio, maximum drawdown, and return over maximum drawdown, which is defined in Section \ref{experiment_stock}.  In this task, we add a new metric:
\begin{itemize}[leftmargin=*]
    \item \textbf{Win/loss ratio} is calculated by dividing the number of winning trades by the number of losing trades.
    %\item \textbf{Win rate and loss rate} calculates the proportion of profitable trades to total trades, and vice versa.
    %providing insight into the strategy’s consistency and reliability.
\end{itemize}

\textbf{Results}. As seen in Table \ref{tab:ensemble_performance_crypto} and Fig. \ref{fig:stock_returns} (b), Double DQN and Dueling DQN agents have similar performance, with cumulative returns of $0.48\%$. This is lower than the BTC price baseline. Despite this, they achieve higher Sharpe ratios of $0.21$ and lower maximum drawdowns of $-0.98\%$ than the fixed-time exit strategy and BTC price baseline, suggesting effective risk management. The three different ensemble models have similar performances, which may be due to the limited action space at each timestep, causing agents to output identical actions. Their cumulative returns are close to the BTC price baseline. Moreover, the ensemble models outperform all individual agents in all metrics, achieving the highest Sharpe ratio of $0.28$ and the lowest maximum drawdown of $-0.73\%$. They also have the highest win/loss ratio of $1.62$. This shows that ensemble methods can mitigate the risks associated with the decision-making failures of single agents.

We observe that individual agents and majority voting ensembles have near identical performances; in the restricted action space, agent policies may be converging indicating a greater risk of significant losses due to less diversified trading actions. Compared to the spot price and fixed-time exit, the ensemble consistently achieved a higher return over the maximum drawdown ratio, highlighting superior risk-adjusted returns. Due to a lack of diversity in agents, we observed near-identical results between ensembles and individual agents. Nonetheless, over a relatively short timespan of $30$ minutes, we observe that the vectorized agents can outperform other strategies.

This was crucial in maintaining portfolio stability amid high market volatility. The ensemble generally showed a higher win rate and a lower loss rate than the individual agents, reflecting its enhanced decision-making accuracy and consistency. The comparative analysis between the ensemble model and individual trading agents underscores the ensemble’s superior capability in managing risks and capitalizing on market opportunities. While individual agents provide valuable insights and are crucial components of the ensemble, the aggregated approach of the ensemble offers a more robust and effective solution for trading in the volatile cryptocurrency markets.

\subsection{Reflection of ACM ICAIF FinRL Contests 2023/2024}

\textbf{Dilemma}: Along the journey of developping the FinRL framework \cite{liu2020finrl,liu2021finrl, liu2022finrl,liu2024dynamic,liu2018practical} and organizing the ACM ICAIF FinRL Contests 2023/2024, we noticed a dilemma: 1). If the obtained FinRL trading agent is powerful, why model producers (or contest participants) do not trade on their own funds privately, instead releasing their methods (and codes, model weights, or other artifacts) to open-source community?  2). Any powerful trading agent may become obsolete inmediately after becoming publicly available.  

\textbf{Plausible solution}: Zero-knowledge proofs (ZKPs) provide a secure protocol between model producers and institutional funds. 
\begin{itemize}[leftmargin=*]
    \item A model producer validates the FinRL trading agent through backtesting or real-time paper trading and generates a proof-file via a ZKP-prover.
    \item A model producer publishes a report or whitepaper (can be anonymous but with contact channels, say emails) that describes the backtesting process or real-time paper trading process (NOT revealing the method and the trained FinRL agent) and publishes the generated proof-file (probably uploads it to a blockchain).
    \item An institutional fund downloads the proof-file and verifies the results in the report via a ZKP-verifier.
    \item The institutional fund may contact the model producer and discuss about collaborations.  1).  The institutional fund could hire the model producer; 2). The two parties could set up a decentralized autonomous organization (DAO) for a new trading strategy, using smart contracts...well, the legend begins!
\end{itemize}

\section{Conclusion}
In this paper, we have revisited ensemble methods and combined them with massively parallel simulation to perform stock and cryptocurrency trading tasks. It enhances the efficiency and robustness of trained models in volatile financial markets. Massively parallel simulations on the GPU improve the sampling speed by up to $1,746\times$ with $2,048$ parallel environments, compared with a single environment. Ensemble methods, combining the strengths of diverse agents to mitigate policy instability, and improve model performance. 
Results in stock and cryptocurrency trading tasks show that the ensemble models have high cumulative returns and outperform some individual agents, with a reduction of up to $4.17\%$ in maximum drawdown and an improvement of up to $0.21$ in the Sharpe ratio.

%achieving lower maximum drawdowns and higher Sharpe ratios.
%Results in stock and cryptocurrency trading tasks show that ensemble methods consistently surpassed individual models in return over max drawdown.

In conclusion, integrating ensemble methods with massively parallel simulations on the GPU is a powerful approach to addressing the policy instability and sampling bottleneck in FinRL. With high stability, ensemble models can achieve high generalization capability. Future research can focus on optimizing these techniques and exploring a broader array of financial instruments and market conditions. Large-scale ensemble collections that benefit from accelerated sampling can combine various agents, leading to more generally capable models in FinRL.

%This may be especially promising when creating large collections of ensembles for multi-task financial agents that benefit from faster sampling. This study lays a foundational pathway for developing more reliable and efficient systems in the evolving landscape of financial technology.  

\begin{acks}

Nikolaus Holzer, Keyi Wang and Xiao-Yang Liu Yanglet acknowledge the support from Columbia's SIRS and STAR Program, The Tang Family Fund for Research Innovations in FinTech, Engineering, and Business Operations.

\end{acks}

\textbf{Disclaimer: We are sharing views and codes for academic purposes under the MIT education license. Nothing herein is financial advice, and NOT a recommendation to trade real money. Please use common sense and always first consult a professional before trading or investing.}

\bibliographystyle{acm_ref_format}
\bibliography{bibliography}


\begin{thebibliography}{36}

%%% the Web address.

\ifx \showCODEN    \undefined \def \showCODEN     #1{\unskip}     \fi
\ifx \showDOI      \undefined \def \showDOI       #1{#1}\fi
\ifx \showISBNx    \undefined \def \showISBNx     #1{\unskip}     \fi
\ifx \showISBNxiii \undefined \def \showISBNxiii  #1{\unskip}     \fi
\ifx \showISSN     \undefined \def \showISSN      #1{\unskip}     \fi
\ifx \showLCCN     \undefined \def \showLCCN      #1{\unskip}     \fi
\ifx \shownote     \undefined \def \shownote      #1{#1}          \fi
\ifx \showarticletitle \undefined \def \showarticletitle #1{#1}   \fi
\ifx \showURL      \undefined \def \showURL       {\relax}        \fi
% invisible to TeX
\providecommand\bibfield[2]{#2}
\providecommand\bibinfo[2]{#2}
\providecommand\natexlab[1]{#1}
\providecommand\showeprint[2][]{arXiv:#2}

\bibitem[Amrouni et~al\mbox{.}(2022)]%
        {amrouni2022abides}
\bibfield{author}{\bibinfo{person}{Selim Amrouni}, \bibinfo{person}{Aymeric Moulin}, \bibinfo{person}{Jared Vann}, \bibinfo{person}{Svitlana Vyetrenko}, \bibinfo{person}{Tucker Balch}, {and} \bibinfo{person}{Manuela Veloso}.} \bibinfo{year}{2022}\natexlab{}.
\newblock \showarticletitle{ABIDES-gym: gym environments for multi-agent discrete event simulation and application to financial markets}. In \bibinfo{booktitle}{\emph{ACM International Conference on AI in Finance}} \emph{(\bibinfo{series}{ICAIF '21})}. \bibinfo{address}{New York, NY, USA}.
\newblock
\showISBNx{9781450391481}


\bibitem[Breiman(1996)]%
        {breiman1996bagging}
\bibfield{author}{\bibinfo{person}{Leo Breiman}.} \bibinfo{year}{1996}\natexlab{}.
\newblock \showarticletitle{Bagging predictors}.
\newblock \bibinfo{journal}{\emph{Machine Learning}} \bibinfo{volume}{24}, \bibinfo{number}{2} (\bibinfo{year}{1996}), \bibinfo{pages}{123--140}.
\newblock


\bibitem[Brockman et~al\mbox{.}(2016)]%
        {brockman2016openaigym}
\bibfield{author}{\bibinfo{person}{Greg Brockman}, \bibinfo{person}{Vicki Cheung}, \bibinfo{person}{Ludwig Pettersson}, \bibinfo{person}{Jonas Schneider}, \bibinfo{person}{John Schulman}, \bibinfo{person}{Jie Tang}, {and} \bibinfo{person}{Wojciech Zaremba}.} \bibinfo{year}{2016}\natexlab{}.
\newblock \showarticletitle{OpenAI gym}.
\newblock \bibinfo{journal}{\emph{arXiv preprint arXiv:1606.01540}} (\bibinfo{year}{2016}).
\newblock


\bibitem[Chan et~al\mbox{.}(2020)]%
        {rl_reliability_metrics}
\bibfield{author}{\bibinfo{person}{Stephanie~CY Chan}, \bibinfo{person}{Samuel Fishman}, \bibinfo{person}{Anoop Korattikara}, \bibinfo{person}{John Canny}, {and} \bibinfo{person}{Sergio Guadarrama}.} \bibinfo{year}{2020}\natexlab{}.
\newblock \showarticletitle{Measuring the Reliability of Reinforcement Learning Algorithms}.
\newblock \bibinfo{journal}{\emph{International Conference on Learning Representations}} (\bibinfo{year}{2020}).
\newblock


\bibitem[Freund and Schapire(1996)]%
        {freund1996boosting}
\bibfield{author}{\bibinfo{person}{Yoav Freund} {and} \bibinfo{person}{Robert~E. Schapire}.} \bibinfo{year}{1996}\natexlab{}.
\newblock \showarticletitle{Experiments with a new boosting algorithm}. In \bibinfo{booktitle}{\emph{Proceedings of the Thirteenth International Conference on International Conference on Machine Learning}} \emph{(\bibinfo{series}{ICML'96})}. \bibinfo{address}{San Francisco, CA, USA}, \bibinfo{pages}{148–156}.
\newblock


\bibitem[Frey et~al\mbox{.}(2023)]%
        {jaxlob}
\bibfield{author}{\bibinfo{person}{Sascha~Yves Frey}, \bibinfo{person}{Kang Li}, \bibinfo{person}{Peer Nagy}, \bibinfo{person}{Silvia Sapora}, \bibinfo{person}{Christopher Lu}, \bibinfo{person}{Stefan Zohren}, \bibinfo{person}{Jakob Foerster}, {and} \bibinfo{person}{Anisoara Calinescu}.} \bibinfo{year}{2023}\natexlab{}.
\newblock \showarticletitle{JAX-LOB: a GPU-accelerated limit order book simulator to unlock large scale reinforcement learning for trading}. In \bibinfo{booktitle}{\emph{ACM International Conference on AI in Finance}} \emph{(\bibinfo{series}{ICAIF '23})}. \bibinfo{address}{New York, NY, USA}, \bibinfo{pages}{583–591}.
\newblock


\bibitem[Fujimoto et~al\mbox{.}(2018)]%
        {scott2018}
\bibfield{author}{\bibinfo{person}{Scott Fujimoto}, \bibinfo{person}{Herke van Hoof}, {and} \bibinfo{person}{David Meger}.} \bibinfo{year}{2018}\natexlab{}.
\newblock \showarticletitle{Addressing function approximation error in actor-critic methods}. In \bibinfo{booktitle}{\emph{International Conference on Machine Learning}}.
\newblock


\bibitem[Ganaie et~al\mbox{.}(2022)]%
        {ensemble_ganaie}
\bibfield{author}{\bibinfo{person}{M.A. Ganaie}, \bibinfo{person}{Minghui Hu}, \bibinfo{person}{A.K. Malik}, \bibinfo{person}{M. Tanveer}, {and} \bibinfo{person}{P.N. Suganthan}.} \bibinfo{year}{2022}\natexlab{}.
\newblock \showarticletitle{Ensemble deep learning: a review}.
\newblock \bibinfo{journal}{\emph{Eng. Appl. Artif. Intell.}} \bibinfo{volume}{115}, \bibinfo{number}{C} (\bibinfo{year}{2022}), \bibinfo{numpages}{18}~pages.
\newblock


\bibitem[Gu et~al\mbox{.}(2023)]%
        {gu2023portfolio}
\bibfield{author}{\bibinfo{person}{Jingyi Gu}, \bibinfo{person}{Wenlu Du}, \bibinfo{person}{A~M~Muntasir Rahman}, {and} \bibinfo{person}{Guiling Wang}.} \bibinfo{year}{2023}\natexlab{}.
\newblock \showarticletitle{Margin trader: a reinforcement learning framework for portfolio management with margin and constraints}. In \bibinfo{booktitle}{\emph{ACM International Conference on AI in Finance}} \emph{(\bibinfo{series}{ICAIF '23})}. \bibinfo{address}{New York, NY, USA}, \bibinfo{numpages}{9}~pages.
\newblock


\bibitem[Haarnoja et~al\mbox{.}(2018)]%
        {haarnoja2018sac}
\bibfield{author}{\bibinfo{person}{Tuomas Haarnoja}, \bibinfo{person}{Aurick Zhou}, \bibinfo{person}{Pieter Abbeel}, {and} \bibinfo{person}{Sergey Levine}.} \bibinfo{year}{2018}\natexlab{}.
\newblock \showarticletitle{Soft actor-critic: off-policy maximum entropy deep reinforcement learning with a stochastic actor}. In \bibinfo{booktitle}{\emph{International Conference on Machine Learning}}, Vol.~\bibinfo{volume}{80}. \bibinfo{publisher}{PMLR}, \bibinfo{pages}{1861--1870}.
\newblock


\bibitem[Hambly et~al\mbox{.}(2023)]%
        {hambly2021recent}
\bibfield{author}{\bibinfo{person}{Ben Hambly}, \bibinfo{person}{Renyuan Xu}, {and} \bibinfo{person}{Huining Yang}.} \bibinfo{year}{2023}\natexlab{}.
\newblock \showarticletitle{Recent advances in reinforcement learning in finance}.
\newblock \bibinfo{journal}{\emph{Mathematical Finance}} \bibinfo{volume}{33}, \bibinfo{number}{3} (\bibinfo{year}{2023}), \bibinfo{pages}{437--503}.
\newblock


\bibitem[Hasselt et~al\mbox{.}(2016)]%
        {doubledqn}
\bibfield{author}{\bibinfo{person}{Hado~van Hasselt}, \bibinfo{person}{Arthur Guez}, {and} \bibinfo{person}{David Silver}.} \bibinfo{year}{2016}\natexlab{}.
\newblock \showarticletitle{Deep reinforcement learning with double Q-Learning}. In \bibinfo{booktitle}{\emph{AAAI Conference on Artificial Intelligence}} \emph{(\bibinfo{series}{AAAI'16})}. \bibinfo{pages}{2094–2100}.
\newblock


\bibitem[Henderson et~al\mbox{.}(2018)]%
        {henderson2018deep}
\bibfield{author}{\bibinfo{person}{Peter Henderson}, \bibinfo{person}{Riashat Islam}, \bibinfo{person}{Philip Bachman}, \bibinfo{person}{Joelle Pineau}, \bibinfo{person}{Doina Precup}, {and} \bibinfo{person}{David Meger}.} \bibinfo{year}{2018}\natexlab{}.
\newblock \showarticletitle{Deep reinforcement learning that matters}. In \bibinfo{booktitle}{\emph{AAAI Conference on Artificial Intelligence}} \emph{(\bibinfo{series}{AAAI'18})}. \bibinfo{numpages}{8}~pages.
\newblock


\bibitem[Jing and Kang(2024)]%
        {jing2024crypto}
\bibfield{author}{\bibinfo{person}{Liu Jing} {and} \bibinfo{person}{Yuncheol Kang}.} \bibinfo{year}{2024}\natexlab{}.
\newblock \showarticletitle{Automated cryptocurrency trading approach using ensemble deep reinforcement learning: Learn to understand candlesticks}.
\newblock \bibinfo{journal}{\emph{Expert Syst. Appl.}}  \bibinfo{volume}{237} (\bibinfo{year}{2024}), \bibinfo{numpages}{20}~pages.
\newblock


\bibitem[John et~al\mbox{.}(2017)]%
        {ppo}
\bibfield{author}{\bibinfo{person}{Schulman John}, \bibinfo{person}{Wolski Filip}, \bibinfo{person}{Dhariwal Prafulla}, \bibinfo{person}{Radford Alec}, {and} \bibinfo{person}{Klimov Oleg}.} \bibinfo{year}{2017}\natexlab{}.
\newblock \showarticletitle{Proximal policy optimization algorithms}.
\newblock \bibinfo{journal}{\emph{arXiv preprint arXiv:1707.06347}} (\bibinfo{year}{2017}).
\newblock


\bibitem[Kakushadze(2016)]%
        {alpha101}
\bibfield{author}{\bibinfo{person}{Zura Kakushadze}.} \bibinfo{year}{2016}\natexlab{}.
\newblock \showarticletitle{101 formulaic alphas}.
\newblock \bibinfo{journal}{\emph{arXiv preprint arXiv:1601.00991}} (\bibinfo{year}{2016}).
\newblock


\bibitem[Lillicrap et~al\mbox{.}(2016)]%
        {ddpg}
\bibfield{author}{\bibinfo{person}{Timothy~P. Lillicrap}, \bibinfo{person}{Jonathan~J. Hunt}, \bibinfo{person}{Alexander Pritzel}, \bibinfo{person}{Nicolas Heess}, \bibinfo{person}{Tom Erez}, \bibinfo{person}{Yuval Tassa}, \bibinfo{person}{David Silver}, {and} \bibinfo{person}{Daan Wierstra}.} \bibinfo{year}{2016}\natexlab{}.
\newblock \showarticletitle{Continuous control with deep reinforcement learning}. In \bibinfo{booktitle}{\emph{International Conference on Learning Representations, {ICLR}}}.
\newblock


\bibitem[Liu et~al\mbox{.}(2022a)]%
        {liu2022finrl}
\bibfield{author}{\bibinfo{person}{Xiao-Yang Liu}, \bibinfo{person}{Ziyi Xia}, \bibinfo{person}{Jingyang Rui}, \bibinfo{person}{Jiechao Gao}, \bibinfo{person}{Hongyang Yang}, \bibinfo{person}{Ming Zhu}, \bibinfo{person}{Christina Wang}, \bibinfo{person}{Zhaoran Wang}, {and} \bibinfo{person}{Jian Guo}.} \bibinfo{year}{2022}\natexlab{a}.
\newblock \showarticletitle{{FinRL-Meta}: Market environments and benchmarks for data-driven financial reinforcement learning}.
\newblock \bibinfo{journal}{\emph{Advances in Neural Information Processing Systems (NeurIPS)}}  \bibinfo{volume}{35} (\bibinfo{year}{2022}), \bibinfo{pages}{1835--1849}.
\newblock


\bibitem[Liu et~al\mbox{.}(2024)]%
        {liu2024dynamic}
\bibfield{author}{\bibinfo{person}{Xiao-Yang Liu}, \bibinfo{person}{Ziyi Xia}, \bibinfo{person}{Hongyang Yang}, \bibinfo{person}{Jiechao Gao}, \bibinfo{person}{Daochen Zha}, \bibinfo{person}{Ming Zhu}, \bibinfo{person}{Christina~Dan Wang}, \bibinfo{person}{Zhaoran Wang}, {and} \bibinfo{person}{Jian Guo}.} \bibinfo{year}{2024}\natexlab{}.
\newblock \showarticletitle{Dynamic datasets and market environments for financial reinforcement learning}.
\newblock \bibinfo{journal}{\emph{Machine Learning - Nature}} (\bibinfo{year}{2024}).
\newblock


\bibitem[Liu et~al\mbox{.}(2018)]%
        {liu2018practical}
\bibfield{author}{\bibinfo{person}{Xiao-Yang Liu}, \bibinfo{person}{Zhuoran Xiong}, \bibinfo{person}{Shan Zhong}, \bibinfo{person}{Hongyang Yang}, {and} \bibinfo{person}{Anwar Walid}.} \bibinfo{year}{2018}\natexlab{}.
\newblock \showarticletitle{Practical deep reinforcement learning approach for stock trading}.
\newblock \bibinfo{journal}{\emph{NeurIPS Workshop on Challenges and Opportunities for AI in Financial Services}} (\bibinfo{year}{2018}).
\newblock


\bibitem[Liu et~al\mbox{.}(2020)]%
        {liu2020finrl}
\bibfield{author}{\bibinfo{person}{Xiao-Yang Liu}, \bibinfo{person}{Hongyang Yang}, \bibinfo{person}{Qian Chen}, \bibinfo{person}{Runjia Zhang}, \bibinfo{person}{Liuqing Yang}, \bibinfo{person}{Bowen Xiao}, {and} \bibinfo{person}{Christina~Dan Wang}.} \bibinfo{year}{2020}\natexlab{}.
\newblock \showarticletitle{{FinRL}: a deep reinforcement learning library for automated stock trading in quantitative finance}.
\newblock \bibinfo{journal}{\emph{Deep RL Workshop, NeurIPS}} (\bibinfo{year}{2020}).
\newblock


\bibitem[Liu et~al\mbox{.}(2022b)]%
        {liu2021finrl}
\bibfield{author}{\bibinfo{person}{Xiao-Yang Liu}, \bibinfo{person}{Hongyang Yang}, \bibinfo{person}{Jiechao Gao}, {and} \bibinfo{person}{Christina~Dan Wang}.} \bibinfo{year}{2022}\natexlab{b}.
\newblock \showarticletitle{FinRL: deep reinforcement learning framework to automate trading in quantitative finance}.
\newblock \bibinfo{journal}{\emph{ACM International Conference on AI in Finance}} (\bibinfo{year}{2022}).
\newblock


\bibitem[Makoviychuk et~al\mbox{.}(2021)]%
        {makoviychuk2021isaac}
\bibfield{author}{\bibinfo{person}{Viktor Makoviychuk}, \bibinfo{person}{Lukasz Wawrzyniak}, \bibinfo{person}{Yunrong Guo}, \bibinfo{person}{Michelle Lu}, \bibinfo{person}{Kier Storey}, \bibinfo{person}{Miles Macklin}, \bibinfo{person}{David Hoeller}, \bibinfo{person}{Nikita Rudin}, \bibinfo{person}{Arthur Allshire}, \bibinfo{person}{Ankur Handa}, {and} \bibinfo{person}{Gavriel State}.} \bibinfo{year}{2021}\natexlab{}.
\newblock \showarticletitle{Isaac gym: high performance GPU based physics simulation for robot learning}. In \bibinfo{booktitle}{\emph{Proceedings of the Neural Information Processing Systems Track on Datasets and Benchmarks}}.
\newblock


\bibitem[Mnih et~al\mbox{.}(2015)]%
        {Mnih2015}
\bibfield{author}{\bibinfo{person}{Volodymyr Mnih}, \bibinfo{person}{Koray Kavukcuoglu}, \bibinfo{person}{David Silver}, {et~al\mbox{.}}} \bibinfo{year}{2015}\natexlab{}.
\newblock \showarticletitle{Human-level control through deep reinforcement learning}.
\newblock \bibinfo{journal}{\emph{Nature}}  \bibinfo{volume}{518} (\bibinfo{year}{2015}), \bibinfo{pages}{529--533}.
\newblock


\bibitem[Mohamed et~al\mbox{.}(2020)]%
        {monte_carlo}
\bibfield{author}{\bibinfo{person}{Shakir Mohamed}, \bibinfo{person}{Mihaela Rosca}, \bibinfo{person}{Michael Figurnov}, {and} \bibinfo{person}{Andriy Mnih}.} \bibinfo{year}{2020}\natexlab{}.
\newblock \showarticletitle{Monte Carlo gradient estimation in machine learning}.
\newblock \bibinfo{journal}{\emph{J. Mach. Learn. Res.}} \bibinfo{volume}{21}, \bibinfo{number}{1} (\bibinfo{year}{2020}), \bibinfo{numpages}{62}~pages.
\newblock


\bibitem[Perez-Cruz(2008)]%
        {perez2008kldivergence}
\bibfield{author}{\bibinfo{person}{Fernando Perez-Cruz}.} \bibinfo{year}{2008}\natexlab{}.
\newblock \showarticletitle{Kullback-Leibler divergence estimation of continuous distributions}. In \bibinfo{booktitle}{\emph{IEEE International Symposium on Information Theory}}. \bibinfo{pages}{1666--1670}.
\newblock


\bibitem[Peters and Bagnell(2010)]%
        {peters2010gradient}
\bibfield{author}{\bibinfo{person}{Jan Peters} {and} \bibinfo{person}{J.~Andrew Bagnell}.} \bibinfo{year}{2010}\natexlab{}.
\newblock \bibinfo{booktitle}{\emph{Policy gradient methods}}.
\newblock \bibinfo{publisher}{Springer US}, \bibinfo{address}{Boston, MA}, \bibinfo{pages}{774--776}.
\newblock


\bibitem[Rafailov et~al\mbox{.}(2023)]%
        {rafailov2023direct}
\bibfield{author}{\bibinfo{person}{Rafael Rafailov}, \bibinfo{person}{Archit Sharma}, \bibinfo{person}{Eric Mitchell}, \bibinfo{person}{Christopher~D Manning}, \bibinfo{person}{Stefano Ermon}, {and} \bibinfo{person}{Chelsea Finn}.} \bibinfo{year}{2023}\natexlab{}.
\newblock \showarticletitle{Direct preference optimization: your language model is secretly a reward model}. In \bibinfo{booktitle}{\emph{Advances in Neural Information Processing Systems}}, Vol.~\bibinfo{volume}{36}. \bibinfo{pages}{53728--53741}.
\newblock


\bibitem[Rudin et~al\mbox{.}(2022)]%
        {rudin2022walk}
\bibfield{author}{\bibinfo{person}{Nikita Rudin}, \bibinfo{person}{David Hoeller}, \bibinfo{person}{Philipp Reist}, {and} \bibinfo{person}{Marco Hutter}.} \bibinfo{year}{2022}\natexlab{}.
\newblock \showarticletitle{Learning to walk in minutes using massively parallel deep reinforcement learning}.
\newblock \bibinfo{journal}{\emph{Proceedings of the 5th Conference on Robot Learning}} (\bibinfo{year}{2022}), \bibinfo{pages}{91--100}.
\newblock


\bibitem[Sutton et~al\mbox{.}(1999)]%
        {sutton1999policy}
\bibfield{author}{\bibinfo{person}{Richard~S Sutton}, \bibinfo{person}{David McAllester}, \bibinfo{person}{Satinder Singh}, {and} \bibinfo{person}{Yishay Mansour}.} \bibinfo{year}{1999}\natexlab{}.
\newblock \showarticletitle{Policy gradient methods for reinforcement learning with function approximation}. In \bibinfo{booktitle}{\emph{Advances in Neural Information Processing Systems}}, Vol.~\bibinfo{volume}{12}.
\newblock


\bibitem[Vittori et~al\mbox{.}(2021)]%
        {vittori2021option}
\bibfield{author}{\bibinfo{person}{Edoardo Vittori}, \bibinfo{person}{Michele Trapletti}, {and} \bibinfo{person}{Marcello Restelli}.} \bibinfo{year}{2021}\natexlab{}.
\newblock \showarticletitle{Option hedging with risk averse reinforcement learning}. In \bibinfo{booktitle}{\emph{ACM International Conference on AI in Finance}} \emph{(\bibinfo{series}{ICAIF '20})}. \bibinfo{address}{New York, NY, USA}, \bibinfo{numpages}{8}~pages.
\newblock


\bibitem[Wang et~al\mbox{.}(2016)]%
        {duelingdqn}
\bibfield{author}{\bibinfo{person}{Ziyu Wang}, \bibinfo{person}{Tom Schaul}, \bibinfo{person}{Matteo Hessel}, \bibinfo{person}{Hado Van~Hasselt}, \bibinfo{person}{Marc Lanctot}, {and} \bibinfo{person}{Nando De~Freitas}.} \bibinfo{year}{2016}\natexlab{}.
\newblock \showarticletitle{Dueling network architectures for deep reinforcement learning}. In \bibinfo{booktitle}{\emph{International Conference on International Conference on Machine Learning}} \emph{(\bibinfo{series}{ICML'16}, Vol.~\bibinfo{volume}{48})}. \bibinfo{pages}{1995–2003}.
\newblock


\bibitem[Wiering and van Hasselt(2008)]%
        {wiering2008ensemble}
\bibfield{author}{\bibinfo{person}{M.~A. Wiering} {and} \bibinfo{person}{H. van Hasselt}.} \bibinfo{year}{2008}\natexlab{}.
\newblock \showarticletitle{Ensemble algorithms in reinforcement learning}.
\newblock \bibinfo{journal}{\emph{Trans. Sys. Man Cyber. Part B}} \bibinfo{volume}{38}, \bibinfo{number}{4} (\bibinfo{year}{2008}), \bibinfo{pages}{930–936}.
\newblock


\bibitem[Yang et~al\mbox{.}(2021)]%
        {yang2020deep}
\bibfield{author}{\bibinfo{person}{Hongyang Yang}, \bibinfo{person}{Xiao-Yang Liu}, \bibinfo{person}{Shan Zhong}, {and} \bibinfo{person}{Anwar Walid}.} \bibinfo{year}{2021}\natexlab{}.
\newblock \showarticletitle{Deep reinforcement learning for automated stock trading: an ensemble strategy}. In \bibinfo{booktitle}{\emph{ACM International Conference on AI in Finance}} \emph{(\bibinfo{series}{ICAIF '20})}. \bibinfo{address}{New York, NY, USA}.
\newblock


\bibitem[Zhang et~al\mbox{.}(2019)]%
        {zhang2019deep}
\bibfield{author}{\bibinfo{person}{Zihao Zhang}, \bibinfo{person}{Stefan Zohren}, {and} \bibinfo{person}{Stephen Roberts}.} \bibinfo{year}{2019}\natexlab{}.
\newblock \showarticletitle{Deep reinforcement learning for trading}.
\newblock \bibinfo{journal}{\emph{Journal of Financial Data Science}} (\bibinfo{year}{2019}).
\newblock


\bibitem[Zhao et~al\mbox{.}(2020)]%
        {zhao2020sim}
\bibfield{author}{\bibinfo{person}{Wenshuai Zhao}, \bibinfo{person}{Jorge~Peña Queralta}, {and} \bibinfo{person}{Tomi Westerlund}.} \bibinfo{year}{2020}\natexlab{}.
\newblock \showarticletitle{Sim-to-real transfer in deep reinforcement learning for robotics: a survey}. In \bibinfo{booktitle}{\emph{IEEE Symposium Series on Computational Intelligence (SSCI)}}. \bibinfo{pages}{737--744}.
\newblock


\end{thebibliography}

\end{document}